\documentclass[%
 reprint,
 superscriptaddress,
 amsmath,amssymb,
 aps,
 prl,
]{revtex4-2}
\usepackage{graphicx}
\usepackage{dcolumn}
\usepackage{bm}

\usepackage{mathtools}
\usepackage{float}
\usepackage[usenames,dvipsnames]{color}
\usepackage[colorlinks=true, citecolor=blue, linkcolor=blue, urlcolor=blue]{hyperref}
\usepackage{bm}
\usepackage[printonlyused]{acronym}
\usepackage{xcolor}

\renewcommand{\vec}[1]{\pmb{#1}}
\newcommand{\op}[1]{\hat{#1}}
\newcommand{\opd}[1]{\hat{#1}^{\dag}}

\usepackage{lipsum}

\begin{document}

\title{Surface-mediated ultra-strong cavity coupling of two-dimensional itinerant electrons}

\newcommand{\affiliationRWTH}{
Institut f\"ur Theorie der Statistischen Physik, RWTH Aachen University and JARA-Fundamentals of Future Information Technology, 52056 Aachen, Germany
}
\newcommand{\affiliationMPSD}{
Max Planck Institute for the Structure and Dynamics of Matter, Center for Free-Electron Laser Science (CFEL), Luruper Chaussee 149, 22761 Hamburg, Germany
}

\newcommand{\affiliationUMD}{
Joint Quantum Institute, Department of Physics, University of Maryland, College Park, MD 20742, USA
}

\newcommand{\affiliationBremen}{
Institute for Theoretical Physics and Bremen Center for Computational Materials Science, University of Bremen, 28359 Bremen, Germany
}

\newcommand{\affiliationFlatiron}{Center for Computational Quantum Physics (CCQ), The Flatiron Institute, 162 Fifth avenue, New York, NewYork 10010, USA}

\author{Christian J. Eckhardt}
\affiliation{\affiliationMPSD}
\affiliation{\affiliationRWTH}

\author{Andrey Grankin}
\affiliation{\affiliationUMD}

\author{Dante M.~Kennes}
\affiliation{\affiliationRWTH}
\affiliation{\affiliationMPSD}

\author{Michael Ruggenthaler}
\affiliation{\affiliationMPSD}

\author{Angel Rubio}
\affiliation{\affiliationMPSD}
\affiliation{\affiliationFlatiron}

\author{Michael A. Sentef}
\affiliation{\affiliationBremen}
\affiliation{\affiliationMPSD}

\author{Mohammad Hafezi}
\affiliation{\affiliationUMD}

\author{Marios H. Michael}
\email[Correspondance to :]{marios.michael@mpsd.mpg.de}
\affiliation{\affiliationMPSD}

\date{\today}
\begin{abstract}
\noindent 
Engineering phases of matter in cavities requires effective light-matter coupling strengths that are on the same order of magnitude as the bare system energetics, coined the ultra-strong coupling regime.
For models of itinerant electron systems, which do not have discrete energy levels, a clear definition of this regime is outstanding to date.
Here we argue that a change of the electronic mass exceeding $10\%$ of its bare value may serve as such a definition.
We propose a quantitative computational scheme for obtaining the electronic mass in relation to its bare vacuum value and show that coupling to surface polariton modes can induce such mass changes.
Our results have important implications for cavity design principles that enable the engineering of electronic properties with quantum light.
\end{abstract}

\maketitle

\paragraph*{Introduction.--}\label{sec:introduction}
Engineering the electromagnetic field inside optical resonators is emerging as a versatile tool to control materials properties \cite{schlawin_cavity_2022, frisk_kockum_ultrastrong_2019, garcia-vidal_manipulating_2021, bloch_strongly_2022, ruggenthaler_quantum-electrodynamical_2014, hubener_engineering_2021}.
In the corresponding field of cavity materials engineering, several intriguing effects have been demonstrated, such as alteration of the critical temperature of a charge density wave transition \cite{jarc_cavity-mediated_2023} and influence on the topologically protected conductance properties of quantum Hall systems \cite{appugliese_breakdown_2022}.
The strong interaction between matter and photon fields in a cavity \cite{frisk_kockum_ultrastrong_2019, forn-diaz_ultrastrong_2019} further gives rise to
the emerging field of polaritonic chemistry \cite{hutchison_modifying_2012, ebbesen_hybrid_2016, ruggenthaler_quantum-electrodynamical_2018, sidler_perspective_2022, ruggenthaler_understanding_2023, garcia-vidal_manipulating_2021, hirai_molecular_2023, nagarajan_chemistry_2021, flick_atoms_2017, simpkins_mode-specific_2021, svendsen_molecules_2023},
promising to revolutionize the way we perceive materials science \cite{frisk_kockum_ultrastrong_2019, schlawin_cavity_2022}.
Theoretically, the change of the electromagnetic field inside cavities is often understood by considering one or few bosonic modes that have a modified light-matter coupling constant \cite{sentef_cavity_2018, frisk_kockum_ultrastrong_2019, schlawin_cavity_2022, svendsen_theory_2023}.
Such models have been successfully employed to describe situations where a material's resonance is coupled to the resonance of a cavity \cite{schafer_shining_2022, basov_towards_2017, ciuti_cavity-mediated_2021, fregoni_theoretical_2022, rokaj_weakened_2023,kipp2024}. 
If the effective coupling constant exceeds $0.1$ times the energy of the mode's frequency, the system is said to be in the ultra-strong coupling regime \cite{frisk_kockum_ultrastrong_2019}.

For systems of itinerant electrons \cite{rokaj_weakened_2023, rokaj_polaritonic_2022, chakraborty_long-range_2021, gao_photoinduced_2020, schlawin_cavity-mediated_2019, andolina_can_2023, vinas_bostrom_controlling_2023, chiocchetta_cavity-induced_2021, fadler_engineering_2024, masuki_cavity_2023}, like a gas of free electrons \cite{rokaj_free_2022}, where no intrinsic materials resonances are present that naturally \emph{select} a particular cavity mode or energy scale, it is a priori unclear whether such few-mode models with an effectively enhanced light-matter coupling strength are applicable \cite{eckhardt_quantum_2022, passetti_cavity_2023, dmytruk_gauge_2021, li_electromagnetic_2020}.
As a consequence, one may need to consider the full continuum of modes, whose coupling strength is generically given by the fine-structure constant $\alpha$, hindering a straightforward extension of the concept of ultra-strong coupling to this situation.

Here, we make progress in this direction by proposing a computational scheme in which changes to the energetics of 2D electrons are computed, in particular their effective physical mass.
When considering the effects of all electromagnetic modes, the change in the mass of the electron due vacuum fluctuations diverges with the high energy cut-off, even in free space\cite{welakuh_non-perturbative_2023, hiroshima_mass_2007,rokaj_free_2022}. To address this issue, we adopt a perturbative renormalization approach, assuming the electron has the measurable physical mass $m_e$ in free space and then computing the relative mass change between free space and different optical setups. 
We define a $10\%$ change of the electronic mass relative to free space as the onset of ultra-strong coupling.
We find that when the electrons are in the near-field of an interface to an insulator hosting phonon polaritons, the ultra-strong coupling regime may indeed be reached when assuming system parameters from experiment. 
We show that this effect is due to coupling to the surface modes at the interface while the bulk modes are to a good approximation unaltered compared to the free-space vacuum. Furthermore we investigate the far-field to near-field crossover close to a Drude metal.

Our results provide a computational tool to investigate the effects different cavity setups have on itinerant electron systems through the photonic density of states.
In addition, we show that significant changes to the electronic mass, underlying several previous theoretical predictions in the field of cavity engineering of itinerant electrons \cite{eckhardt_quantum_2022, dmytruk_controlling_2022,Jon23, latini_ferroelectric_2021,Hector24, rokaj_free_2022, vlasiuk_cavity-induced_2023}, may indeed be experimentally accessible through cavities employing surface phonon polariton modes.
We make a concrete proposal to experimentally test our predictions.

\paragraph*{Electronic self-energy and partial local density of states.--}\label{sec:mass}
\begin{figure}
    \centering
    \includegraphics{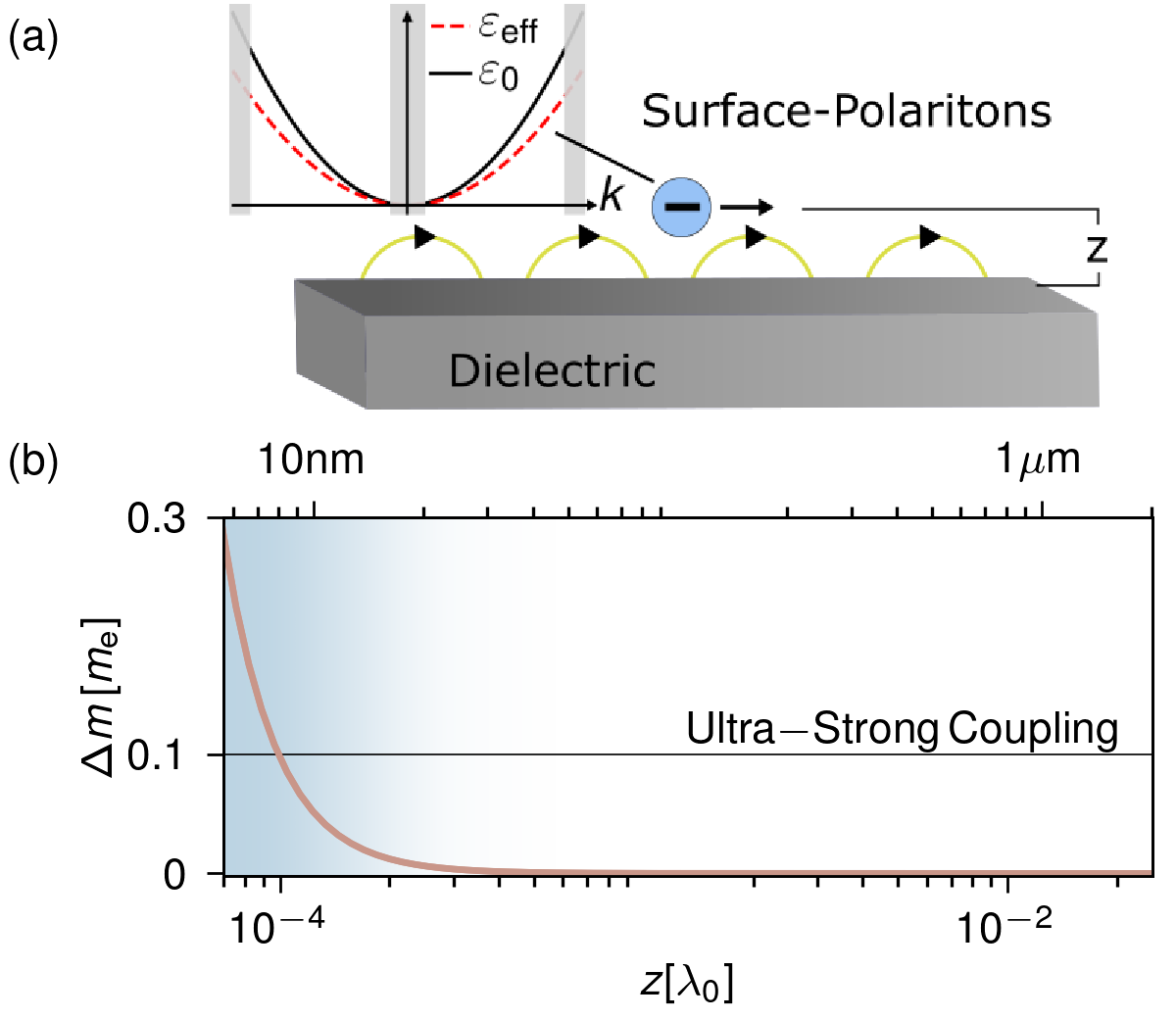}
    \caption{\textbf{Summary of results:}
        \textbf{a.)} Illustration of a 2D electron (blue circle) with quadratic bare dispersion $\varepsilon_0$ (black line) travelling parallel to a surface hosting surface polaritons (yellow semi-circles) at distance $z$. The coupling leads to a flattening of the electronic dispersion $\varepsilon_{\rm eff}$ (red dashed line) corresponding to an effectively increased electronic mass. The $k^2$ dispersion is not valid at infinitely long or short length-scales, marked by the gray bars.
        \textbf{b.)} Change of the mass $\Delta m$ (orange line) as computed via Eq.~(\ref{eq:DeltaM}) in units of the physical electronic mass $m_e$ in vacuum close to an interface with a material that hosts phonon polaritons as a function of distance $z$ to that interface. The onset of strong coupling is marked by the thin black line and blue shading.
        The bottom x-axis is in units of the length scale associated with the surface phonon polaritons $\lambda_0 = \frac{2 \pi c}{\omega_{\infty}}$ where $\sqrt{2}\omega_{\infty} = \sqrt{\omega_{\rm LO}^2 + \omega_{\rm TO}^2}$. Parameters are chosen for the soft phonon of $\mathrm{SrTiO}_3$ -- see main text. 
    }
    \label{fig:1}
\end{figure}
\begin{figure*}
    \centering
    \includegraphics{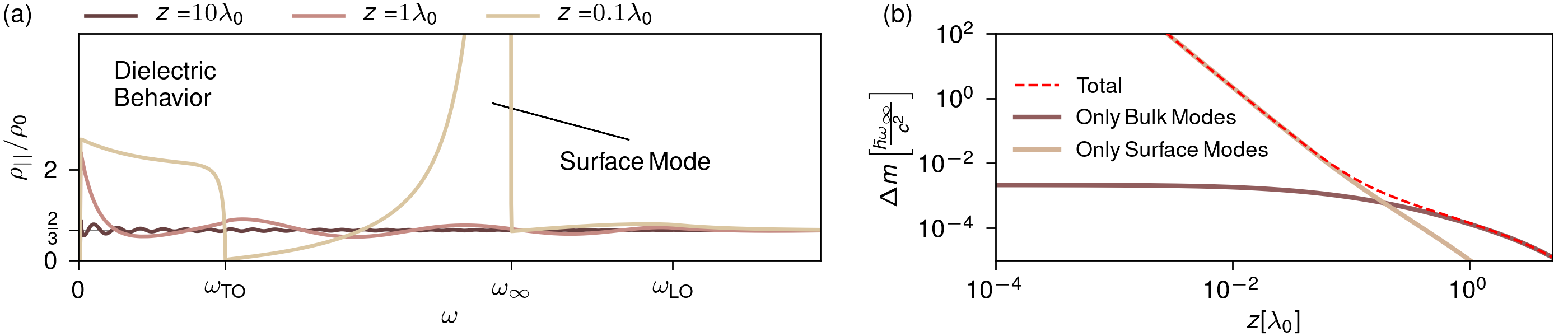}
    \caption{
        \textbf{Mass change due to surface modes}
        \textbf{a.)} In plane partial local density of states, $\rho_{||}$ in units of the free-space photonic density of states $\rho_0 = \frac{\omega^2}{\pi^2 c^3}$ as a function of frequency $\omega$ close to an interface hosting phonon-polaritons for different distances $z$ to the interface (color lines) in units of the length scale associated with the surface phonon polaritons $\lambda_0 = \frac{2 \pi c}{\omega_{\infty}}$ where $ \sqrt{2} \omega_{\infty} = \sqrt{\omega_{\rm LO}^2 + \omega_{\rm TO}^2}$.
        \textbf{b.)} Mass change $\Delta m$ of an electron relative to free-space as a function of the distance to an interface $z_0$ with a material hosting phonon-polaritons only accounting for the surface modes (yellow line, Eq.~(\ref{eq:MassOnlySurfaceModes})), only taking the bulk modes into account (brown line) and the full result (red dashed line). $\omega_{\rm LO}$ and $\omega_{\rm TO}$ are chosen as in Fig.~\ref{fig:1} (b) but the units are chosen such that a global rescaling of frequencies yields the same plot.
    }
    \label{fig:2}
\end{figure*}
In this work we consider the full continuous electromagnetic field close to an interface with a material that we describe using a frequency dependent $\varepsilon(\omega, \vec r)$, see Fig.~\ref{fig:1} (a).
We employ Coulomb gauge and expand the electromagnetic vector potential $\op{\vec A}(\vec r)$ in its eigenmodes:
\begin{equation}
    \vec {\op{A}} (\vec r) = \sum_q \sqrt{\frac{\hbar}{2 \omega_q \varepsilon_0}} e^{-i \vec q_{||} \cdot \vec r_{||}} \vec f_{q_z, \lambda}(z) \left( \opd a_q + \op a_q \right).
    \label{eq:AEigenmodes}
\end{equation}
Here the multi-index $q$ is given as $q = (\vec q_{||}, q_z, \lambda)$ with $\vec q_{||}$ the in-plane part of the wave-vector, $q_z$ its z-component and $\lambda$ denoting the polarization.
$\hbar \omega_q$ is the eigenenergy of the modes where a photon is created by $\opd{a}_q$ and annihilated by $\op{a}_q$.
The mode-functions factorize into a plane-wave part that we have written explicitly and a perpendicular dependence $\vec f_{q_z, \lambda}(z)$ since we assume in-plane translational symmetry.
We provide the explicit mode functions in the Supplementary material 2\cite{supp}.
With these one may compute the partial local density of states (PLDOS) of the electromagnetic field inside the cavity as \cite{barnes_classical_2020}
\begin{equation}
    \rho_{\vec n}(\omega, \vec r) = \sum_q |\vec f_{q_z, \lambda}(\vec r) \cdot \vec n|^2 \delta(\omega - \omega_q),
\label{eq:DefPLDOS}
\end{equation}
where $\vec n$ is a vector of unit length.
In free space the PLDOS is given as \cite{barnes_classical_2020}
\begin{equation}
    \rho_{\vec n, 0}(\omega, \vec r) = \frac{\omega^2}{3 \pi^2 c^3}.
    \label{eq:PLDOSFree}
\end{equation}

We consider a free electron in two dimensions with system-size $L \times L$ that we couple to light in the minimal coupling scheme, assuming consistent in-plane periodic boundary conditions.
The system size sets the longest length-scale at which the $k^2$-dispersion is valid -- also see Fig.~1 (a).
In order to also set a shortest length-scale, we will later introduce a UV cutoff.
We omit spin for brevity and neglect the diamagnetic term that leads to the diamagnetic shift at finite electronic density \cite{rokaj_free_2022}.
Employing the dipole approximation, light and matter can be decoupled exactly by means of a Schrieffer-Wolf type transformation which yields the effective Hamiltonian \cite{eckhardt_quantum_2022, rokaj_free_2022}
\begin{equation}
\begin{aligned}
    &\op H = \sum_q \omega_q \opd a_q \op a_q + \left( \frac{\hbar^2 \vec k_{||}^2}{2 m_0} + \Sigma(\vec k_{||})\right) \opd c_{\vec k_{||}} \op c_{\vec k_{||}}
\end{aligned}
\label{eq:LMCHamiltonian}
\end{equation}
where $\opd c_{\vec k_{||}}$ creates an electron with in-plane momentum $\hbar \vec k_{||}$ while $\op c_{\vec k_{||}}$ annihilates such an electron and $m_0$ is the bare mass. 
We note that beyond the dipole approximation Eq.~(\ref{eq:LMCHamiltonian}) would receive further corrections that may, however, be neglected by perturbative arguments.
In Eq.~(\ref{eq:LMCHamiltonian}), the effect of the photons onto the electrons is incorporated via the self-energy $\Sigma$ that is defined as
\begin{equation}
    \Sigma(\vec k_{||}) = - \sum_{q} \frac{ e^2 \hbar^2 |\vec k_{||} \cdot \vec f_{q_z, \lambda}(z_0)|^2}{m_0^2 \hbar \omega_q} \frac{\hbar}{2 \omega_q \varepsilon_0}.
    \label{eq:SigmaSum}
\end{equation}
This self-energy explicitly depends on the continuous three-dimensional electromagnetic field through the mode functions and can be directly related to the PLDOS of the electromagnetic field using its definition in Eq.~(\ref{eq:DefPLDOS}),
\begin{equation}
    \Sigma(\vec k_{||}, \Lambda_{\rm UV}) = - \alpha \frac{ 4 \pi \hbar^3 c |\vec k_{||}|^2}{2 m_0^2 } \int_{\Lambda_{\rm IR}}^{\Lambda_{\rm UV}} \mathrm{d}\omega \, \frac{\rho_{\vec e_{\vec q_{||}}}(\omega, \vec r)}{\omega^2},
\label{eq:SigmaDos}
\end{equation}
where $\alpha$ is the fine-structure constant. We consider a high-energy cut-off that ensures the validity of the dipole approximation, $\Lambda_{\rm UV} \ll c |k_{\parallel}|$. For typical momenta of interest in condensed matter, $k_{\parallel} \sim 1$\AA, this condition also guarantees that the cut-off we consider is below the Landau pole, $\Lambda_{\rm Landau}$, for which perturbative non-relativistic QED within the dipole approximation becomes ill-defined \cite{hiroshima_mass_2007}.
Additionally, we have included a low-energy cutoff $\Lambda_{\rm IR}$ which is not strictly necessary to regularize non-relativistic QED \cite{hiroshima_mass_2007, welakuh_non-perturbative_2023} but is instead introduced for two reasons: (i) in order to be agnostic to specific boundary conditions that we set for the electromagnetic field discussed later; (ii) to account for the fact that the electronic $k^2$ dispersion is naturally not valid on infinite length scales (see Fig.\ref{fig:1}(a)) which effectively necessitates a low energy cutoff on the combined light-matter system.
The latter point is further illustrated by the fact that in reality, light couples electrons of momentum $k$ to $k + q$, and therefore the IR cut-off of the electronic system also creates a cut-off to the light modes that are able to couple to electrons.
For a gas of free electrons one would equally expect the self-energy to be given by Eq.~(\ref{eq:SigmaDos}) from perturbative arguments (Hartree-Fock) but including an extra phase-factor due to Pauli blocking.
In this case, for a sizeable electronic density, the effect of the diamagnetic term would need to be reconsidered.
\paragraph*{Mass renormalization relative to free space.--}

Incorporating the self-energy into an effective dispersion $\varepsilon_{\rm eff} (\vec k_{||}) = \frac{\hbar^2 \vec k_{||}^2}{2 m_0} + \Sigma(\vec k_{||})$, naturally gives rise to a definition of the effective mass of an electron with an electromagnetic environment as $m_{\rm eff} = \hbar^2 \left[ \partial_{\vec k_{||}}^2 \varepsilon_{\rm eff}(\vec k_{||}) \right]^{-1}$.
Note that due to the dipole approximation as well non-relativistic limit the self-energy is real and frequency independent.
From Eq.~(\ref{eq:SigmaDos}) it becomes clear that the self-energy depends on the UV cut-off and diverges when removing the cut-off $\Lambda_{\rm UV}$ in free space where the PLDOS is given by Eq.~(\ref{eq:PLDOSFree}).
This divergence in QED is usually mitigated perturbatively by using mass renormalization, where the coupling to the electromagnetic field leads to the replacement of a cut-off dependent bare mass $m_0$ by the physical mass $m_{e}$ \cite{schwartz_quantum_2014}:
\begin{equation}
    m_0( \Lambda_{\rm UV})  = m_e + \frac{m_e^2}{\hbar^2}\partial_{\vec k_{\parallel}}^2 \Sigma_{\rm vac}( \vec k_{\parallel} , \Lambda_{ \rm UV}) + \mathcal{O}(\alpha^2),
\label{eq:vac}
\end{equation}
to leading order in the fine-structure constant.
To fix the bare mass, we assume that the effective mass of the electron far away from the interface, where the density of states asymptotically approaches the free space one, Eq.~(\ref{eq:PLDOSFree}), is given by the physical mass $m_e$ for any given cut-off $\Lambda_{\rm UV}$.
This allows us to compute the change in the effective mass relative to this point of reference: 
\begin{equation}
    \Delta m := m_{\rm eff} - m_e = - \frac{m_e^2}{\hbar^2} \partial_{\vec k_{||}}^2\Sigma_{\rm eff}(\vec k_{||}) + \mathcal{O}(\alpha^2),
\label{eq:DeltaM}
\end{equation}
which depends on an effective self-energy in which a reference PLDOS, $\rho_{\vec n, r}$, is subtracted from the cavity one:
\begin{equation}
    \Sigma_{\rm eff}(\vec k_{||}) = - \alpha \frac{4 \pi \hbar^3 c |\vec k_{||}|^2}{2 m_e^2} \int_{\Lambda_{\rm IR}}^{\Lambda_{\rm UV}} \mathrm{d}\omega \, \frac{\rho_{\vec e_{\vec q_{||}}}(\omega, \vec r) - \rho_{\vec n, r}(\omega)}{\omega^2}.
\label{eq:SigmaEff}
\end{equation}
The above expression, Eq.~(\ref{eq:SigmaEff}), quantifies the effective enhancement or reduction of the electron-photon coupling inside a cavity through integrating the difference in the PLDOS compared to free space and constitutes the first major result of our work.
\paragraph*{Redistribution of spectral weight in the PLDOS.--}
\begin{figure*}
    \centering
    \includegraphics{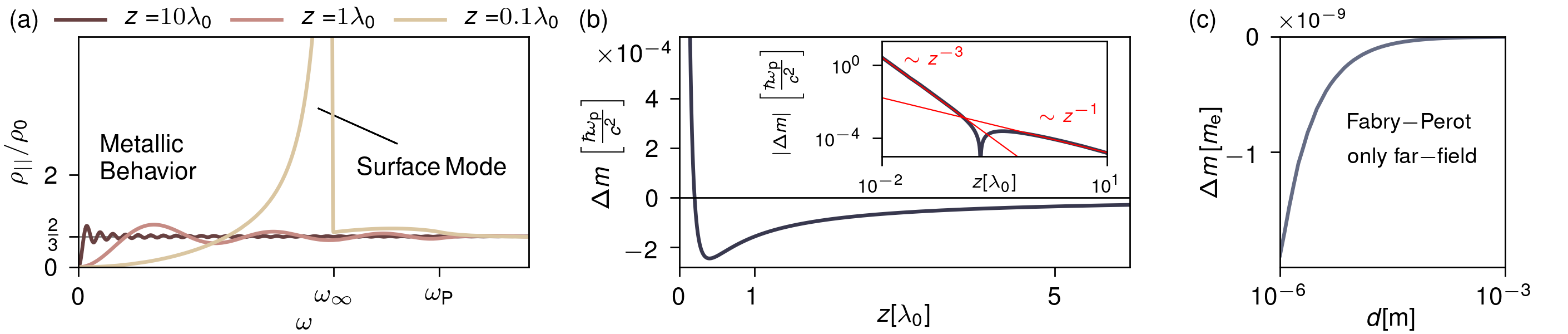}
    \caption{
        \textbf{Mass change close to a Drude metal}
        \textbf{a.)}  In plane partial local density of states $\rho_{||}$ (colored lines) in units of the free-space photonic density of states $\rho_0 = \frac{\omega^2}{\pi^2 c^3}$ as a function of frequency $\omega$ close to a Drude metal for different distances $z$ to the interface (color lines) in units of the length scale associated with the surface plasmon-polaritons $\lambda_0 = \frac{2 \sqrt{2} \pi c}{\omega_{\rm P}}$.
        \textbf{b.)} Mass change $\Delta m$ (blue line) as computed with Eq.~(\ref{eq:DeltaM}) as a function to distance $z$ to a Drude metal in units of $\lambda_0$ defined as in \textbf{a.)}. The units of the plot are chosen such that a rescaling of the plasma-frequency $\omega_{\rm P}$ does not change the plot. In the inset the absolute value of the same quantity is plotted on a log-log scale with a scaling as $z^{-3}$ in the near field and $z^{-1}$ in the far field indicated (red lines).
        \textbf{c.)} Relative mass change of a free electron (blue line) in the middle of an idealized Fabry-P\'erot cavity (perfect metal boundary conditions) assuming that the $k^2$-dispersion of the electron is valid at energy-scales far below the fundamental frequency of the cavity.
    }
    \label{fig:3}
\end{figure*}

We compute the change of mass close to a phonon-polariton system.
We assume the lower half-space to be filled with the insulating material giving rise to a frequency dependent permittivity that we take as the one produced by a Lorentz-oscillator model for the phonons \cite{gubbin_real-space_2016}
\begin{equation}
    \varepsilon(\omega, \vec r) = \begin{cases}
        &1 \hspace{23mm} ; \,z> 0\\
        &\varepsilon(\omega) = \frac{\omega_{\rm LO}^2 - \omega^2}{\omega_{\rm TO}^2 - \omega^2} \hspace{2mm} ; \, z < 0
    \end{cases}
\label{eq:Permittivity}
\end{equation}
where $\omega_{\rm LO}$ is the longitudinal phonon frequency while $\omega_{\rm TO}$ the transversal one, $\omega_{\rm LO} < \omega_{\rm TO}$ \cite{mahan_many-particle_2000}.
In order to obtain concrete numbers we choose $\omega_{\rm LO} = 2 \pi \times 5.1$THz and $\omega_{\rm TO} = 2\pi \times 1.26$THz which corresponds to the soft phonon of $\mathrm{SrTiO}_3$. 
As the point of reference we choose a distance to the interface $z$ for which $z \gg \frac{c}{\omega_{\rm LO}}$, in particular $z = 0.25$m.
In order to obtain discrete modes we assume perfect metal boundary conditions in $z$-direction with a system extent $L_{z} \gg \frac{c}{\omega_{\rm LO}}$, in particular choosing $L_{z} = 1$m, and choose an IR cut-off of $ \Lambda_{\rm IR} = 50$GHz considerably above the corresponding fundamental mode associated with the $1$m length scale.
We have confirmed numerically that any cutoff $\Lambda_{\rm IR}$ far below the energy-scales related to the surface phonon-polariton, in particular $\omega_{\rm TO}$, does not change the results quantitatively.
On the other hand we use a UV cut-off of $\hbar\Lambda_{\rm UV} = 0.3$eV, such that $\Lambda_{\rm UV} \gg \omega_{\rm LO}$. Again we have checked that the results obtained via Eq.~(\ref{eq:DeltaM}) do not change as long as the cutoff is far above the energies related to the surface phonon-polariton, in particular $\omega_{\rm LO}$.
We report the relative change of mass in Fig.\ref{fig:1} (b), which shows that for realistic distances to the surface the change in mass may indeed exceed $10\%$ of the physical mass in free-space.
We coin such a drastic change in mass of a free electron as the ultra-strong coupling regime for this case \cite{frisk_kockum_ultrastrong_2019}. 
Details of the calculation of the PLDOS are given in Supplementary Material 3\cite{supp}. 
We explain the results of the previous part in terms of a redistribution of spectral weight in the PLDOS which we show in Fig.~\ref{fig:2} (a).
At low frequencies, $\omega < \omega_{\rm TO}$, spectral weight is added since the material acts as a dielectric in this regime.
Within the restrahlenband and above, $\omega > \omega_{\rm TO}$, spectral weight is depleted since $\varepsilon(\omega)|_{\omega > \omega_{\rm TO}} < 1$, with some oscillations at high frequencies.
On top of this a strong peak around $\omega \lesssim \omega_{\infty} = \sqrt{\frac{\omega_{\rm LO}^2 + \omega_{\rm TO}^2}{2}}$ appears close to the interface, $z < \frac{c}{\omega_{\infty}}$, that can be attributed to the surface modes.

We compare the mass renormalization induced by surface and bulk modes separately. The result is shown in Fig.~\ref{fig:2} (b) using the same parameters as for Fig.~\ref{fig:1} (b) but choosing natural units such that a global rescaling of all frequencies does not alter the results. 
We find that the bulk mode contribution leads to a mass change only weakly depending on distance that makes up only a negligible portion of the overall mass change close to the interface.
As a consequence, close to the interface, $z < \frac{c}{\omega_{\infty}}$, the bulk mode contribution nearly cancels the free space contribution in Eq.~(\ref{eq:DeltaM}), and the mass renormalization is dominated by the additional presence of the surface modes, i.e.:
\begin{equation}
    \int_{\Lambda_{\rm IR}}^{\Lambda_{\rm UV}} \mathrm{d}\omega \, \frac{\rho_{\vec e_{\vec q_{||}}}(\omega, \vec r) - \rho_{\vec n, r}(\omega)}{\omega^2} \approx \int_{\omega_{\rm TO}}^{\omega_{\rm LO}} \frac{\rho_{\vec e_{\vec q_{\parallel}}}^{\rm SPP}(\omega, \vec r )}{\omega^2},
    \label{eq:MassOnlySurfaceModes}
\end{equation}
where $\rho_{e_{\vec q_{\parallel}}}^{\rm SPP}(\omega, \vec r)$ denotes the density of states of the surface modes only. This allows us to derive an approximate analytic expression for the mass renormalization as the electron approaches the interface (derived in Supplementary Material 4\cite{supp}):
\begin{equation}
    \Delta m_{\rm SPhP} = \alpha \frac{\hbar c (\omega_{\rm \infty}^2 - \omega^2_{\rm TO})}{4 \omega_{\infty}^4 } \frac{1}{z^3}.
    \label{eq:MassChangeDueToSPhP}
\end{equation}
These findings justify a computational approach for quantifying effects from coupling to surface modes where bulk modes only enter implicitly in the calculation giving rise to the bare system parameters while the surface modes are quantized and treated explicitly within the theory \cite{lenk_dynamical_2022, eckhardt_theory_2024}.
\paragraph*{Drude Metal.--}

In this part we consider an electron moving close to a Drude metal. 
We model the Drude metal by a frequency dependent permittivity $\varepsilon(\omega)$ that takes the same form as Eq.~(\ref{eq:Permittivity}) but setting $\omega_{\rm P} \equiv \omega_{\rm LO}$ and $\omega_{\rm TO} = 0$ and assume small losses such that we have for the scattering time $\tau^{-1} \ll \omega_{\rm P}$.
We show the in-plane PLDOS close to the interface in Fig.~\ref{fig:3} (a).
A strong peak appears for small distances $z \ll \frac{c}{\omega_{\rm P}}$ associated with the surface plasmon-polariton mode.
At low frequency the density of states is reduced since the Drude metal acts as a perfect conductor in this regime.

We compute the effective change in mass for different distances to the interface using Eq.~(\ref{eq:DeltaM}).
Again we introduce cutoffs but note that the results don't change quantitatively as long as $\Lambda_{\rm IR} \ll \omega_{\rm P} \ll \Lambda_{\rm UV}$ is fulfilled.
The result is shown in Fig.~\ref{fig:3} (b).
We find two distinct regimes: A far-field regime, $z \gg \frac{2 \pi c}{\omega_{\rm P}}$ where the effective mass of the electrons is decreased scaling as $\Delta m \sim - z^{-1}$; and a near-field regime, $z \ll \frac{2 \pi c}{\omega_{\rm P}}$, where the effective mass increases as $\Delta m \sim z^{-3}$.
The increase close to the interface is due to coupling to surface modes, surface plasmon-polaritons in this case, similar to the case of the insulator presented in Fig.~\ref{fig:2}.
In the far field, we attribute the decrease in mass to the perfect metal properties of the Drude metal at low frequencies signified by the depletion of spectral weight at low frequencies $\omega \ll \omega_{\rm P}$.

The far-field effect of a Drude metal on the electronic mass has interesting implications for electrons in an idealized Fabry-P\'erot cavity.
One may model a Fabry-P\'erot cavity by perfect mirror boundary conditions at $z = 0$ and $z = d$ giving rise to the fundamental frequency $\omega_0 = \frac{\pi c}{d}$.
This reduces the effect of the mirrors to far-field effects of a metal and leads to the gapping out of all in-plane modes below the fundamental frequency $\omega_0$.
We now assume a free electron at $z = \frac{d}{2}$ that has a $k^2$-dispersion which is valid far below the fundamental frequency of the cavity, as well as far above, $\Lambda_{\rm IR} \ll \omega_0 \ll \Lambda_{\rm UV}$. 
We find a decrease of the mass of the electron given by fundamental constants, $\delta m/m_e \sim -\frac{\alpha \hbar}{ c d m_e}$, signifying a reduced effective light-matter coupling compared to free space -- see Fig.~\ref{fig:3}(c).
This change can be attributed to the gapping out of in-plane light-modes below $\omega_0$ to which the electron is sensitive when assuming $\Lambda_{\rm{IR}} \ll \omega_0$. In the case of $\Lambda_{\rm IR} \approx \omega_0$, which could be relevant for bound systems\cite{welakuh_non-perturbative_2023} a different result may be expected.
More details on the calculation for the idealized Fabry-P\'erot cavity can be found in the supplement\cite{supp}. 
\paragraph*{Discussion.--}
We have investigated the coupling of 2D free electrons to light, treating the full continuum of modes of the electromagnetic field instead of relying on single- or few-mode approximations.
We extend the definition of ultra-strong coupling to this case as a $10\%$ change in the effective electronic mass, which we compute using a method that we propose in this work.
Near-field effects close to an insulator can lead to this ultra-strong coupling regime due to coupling to surface modes. 
Importantly we find that one may compute the mass change of the electron solely from the coupling to the surface modes without the need to renormalize with respect to the vacuum \cite{lenk_dynamical_2022, eckhardt_theory_2024}.
These results further emphasize the experimental prospects of engineering electronic properties via cavities employing near-field effects \cite{appugliese_breakdown_2022, kipp_cavity_2024, thomas_exploring_2019, thomas_large_2021, scalari_ultrastrong_2012, maissen_ultrastrong_2014, bayer_terahertz_2017, halbhuber_non-adiabatic_2020, yu_plasmon-enhanced_2019, tame_quantum_2013, riolo_tuning_2024, herzig_sheinfux_high-quality_2024, 2408.00189}.

It is interesting to consider the applicability of sum rules to the change of mass.
A sum rule derived in Ref.~\citenum{barnett_sum_1996} directly applies to a perfect metal reproducing the $\Delta m \sim - z^{-1}$ scaling found in the present work in the far-field of a Drude metal, while close to interfaces sum rules derived for a frequency dependent permittivity $\varepsilon(\omega)$, in Ref.~\citenum{scheel_sum_2008} and Ref.~\citenum{sanders_analysis_2018}, are consistent with the scaling $\Delta m \sim z^{-3}$ found here.

The mass of free electrons directly enters in their cyclotron frequency and is well documented in the literature for many materials. As a precise measurement of the predictions made in this work, we therefore propose to measure the cyclotron resonance in a dilute, two-dimensional gas of electrons close to an insulator that hosts suitable surface phonon polaritons, for example $\mathrm{SrTiO}_3$.

We note that here we focus on the single-particle coupling. Strong coupling effects could be further amplified by collective strong-coupling, where a macroscopic amount of particles couples to the photonic modes, such as in the case of polaritonic chemistry \cite{garcia-vidal_manipulating_2021,Ebbesen23}. Future studies should investigate the complex behavior expected once the inter-particle interactions of a macroscopic ensemble is included\cite{Sidler24,ruggenthaler_understanding_2023,Sidler21}.

\paragraph*{Acknowledgments.--}
We acknowledge useful discussions with Junichiro Kono, Andrey Baydin, Eugene Demler, Jonathan Curtis, and Maximilian Daschner.  M. H. M. is grateful for the support from the Alexander von Humboldt foundation. M. A. S. was funded by the European Union (ERC, CAVMAT, project no. 101124492). D.M.K. acknowledge support by the DFG via Germany’s Excellence Strategy -Cluster of Excellence Matter and Light for Quantum Computing (ML4Q, Project No. EXC 2004/1, Grant No. 390534769), individual grant No. 508440990. A. R. and M. R. acknowledge support from the Cluster of Excellence ``CUI: Advanced Imaging of Matter''- EXC 2056 - project ID 390715994 and SFB-925 ``Light induced dynamics and control of correlated quantum systems'' – project 170620586 of the Deutsche Forschungsgemeinschaft (DFG), and Grupos Consolidados (IT1453-22), and from the Max Planck-New York City Center for Non-Equilibrium Quantum Phenomena at the the Flatiron Institute. The Flatiron Institute is a division of the Simons Foundation. M. H. and A.G. acknowledge support by DARPA. 

\bibliography{CavitronicsPaper}

\end{document}


\title{Supplement for \textit{Surface-mediated ultra-strong cavity coupling of two-dimensional itinerant electrons}}

\newcommand{\affiliationRWTH}{
Institut f\"ur Theorie der Statistischen Physik, RWTH Aachen University and JARA-Fundamentals of Future Information Technology, 52056 Aachen, Germany
}
\newcommand{\affiliationMPSD}{
Max Planck Institute for the Structure and Dynamics of Matter, Center for Free-Electron Laser Science (CFEL), Luruper Chaussee 149, 22761 Hamburg, Germany
}

\newcommand{\affiliationUMD}{
Joint Quantum Institute, Department of Physics, University of Maryland, College Park, MD 20742, USA
}

\newcommand{\affiliationBremen}{
Institute for Theoretical Physics and Bremen Center for Computational Materials Science, University of Bremen, 28359 Bremen, Germany
}

\newcommand{\affiliationFlatiron}{Center for Computational Quantum Physics (CCQ), The Flatiron Institute, 162 Fifth avenue, New York, NewYork 10010, USA}

\author{Christian J. Eckhardt}
\affiliation{\affiliationMPSD}
\affiliation{\affiliationRWTH}

\author{Andrey Grankin}
\affiliation{\affiliationUMD}

\author{Dante M.~Kennes}
\affiliation{\affiliationRWTH}
\affiliation{\affiliationMPSD}

\author{Michael Ruggenthaler}
\affiliation{\affiliationMPSD}

\author{Angel Rubio}
\affiliation{\affiliationMPSD}
\affiliation{\affiliationFlatiron}

\author{Michael A. Sentef}
\affiliation{\affiliationBremen}
\affiliation{\affiliationMPSD}

\author{Mohammad Hafezi}
\affiliation{\affiliationUMD}

\author{Marios H. Michael}
\email[Correspondance to :]{marios.michael@mpsd.mpg.de}
\affiliation{\affiliationMPSD}

\date{\today}

\maketitle
\renewcommand{\theequation}{S.\arabic{equation}}
\renewcommand\thefigure{S.\arabic{figure}} 
\subsection*{Supplement 1: Details on the calculations of effective mass in an idealized Fabry-P\'erot cavity}
\begin{figure}
    \centering
    \includegraphics{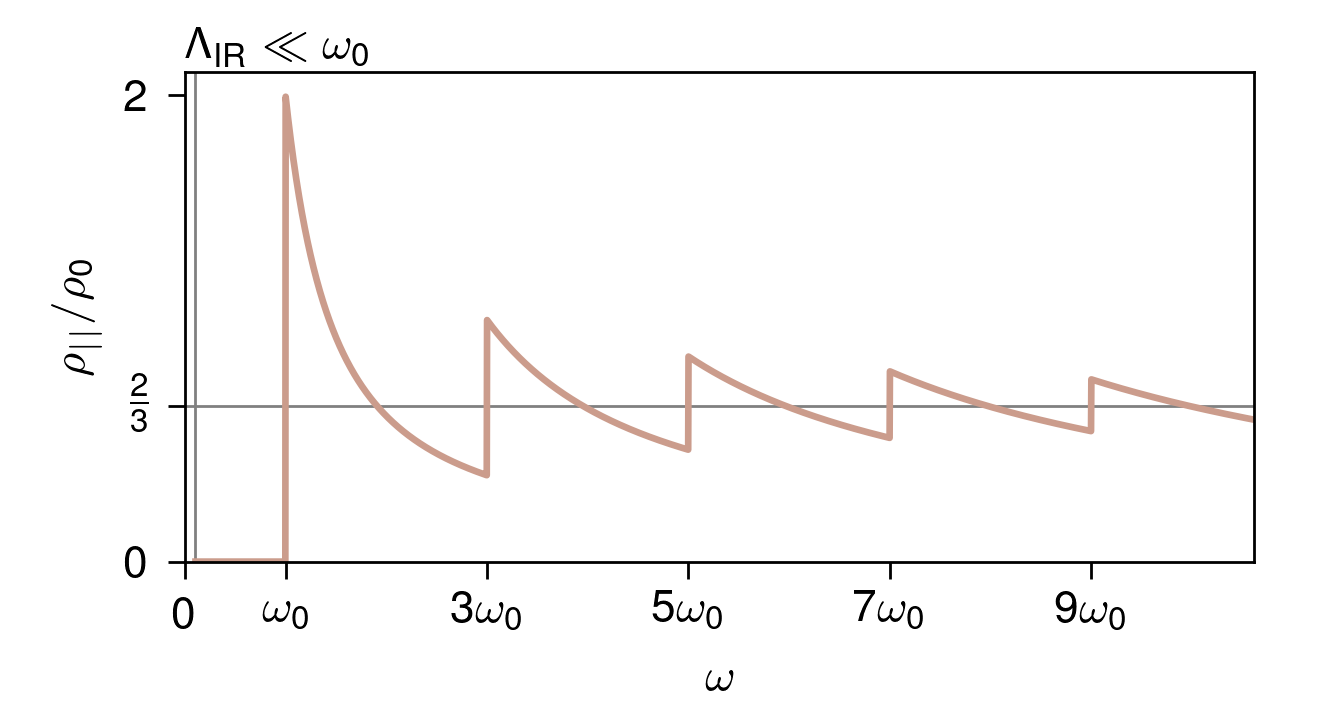}
    \caption{\textbf{Density of states of an idealized Fabry-P\'erot cavity.}
    In plane partial local density of states $\rho_{||}$ (orange line) in units of the free-space photonic density of states $\rho_0 = \frac{\omega^2}{\pi^2 c^3}$ as a function of frequency $\omega$ inside an idealized Fabry-P\'erot cavity with mirror separation $d$ at $z = \frac{d}{2}$.
    Due to assuming perfect mirror boundary conditions and an infinitely extended cavity, all in-plane modes below $\omega_0$ are gapped out. We have schematically marked the low-energy cutoff $\Lambda_{\rm IR}$ that, in our calculation, is set significantly below the fundamental resonance.}
    \label{fig:s1}
\end{figure}
We model the idealized Fabry-P\'erot cavity, used to obtain Fig.~3 (c) in the main part, by perfect mirror boundary conditions at $z = 0$ and $z = d$.
The resulting in-plane PLDOS at $z = \frac{d}{2}$ is shown in Fig.~\ref{fig:s1}.
In particular, all in-plane modes below $\omega_0 = \frac{\pi c}{d}$ are gapped out due to the assumed geometry of the cavity (infinite mirror in the $xy$-plane) and due to neglecting losses.
A realistic cavity is expected to not influence the electromagnetic field at infinite length scales such that one expects the local DOS to approach its free-space value towards $\omega \to 0$.
The precise form of the mode-functions and details on the calculation of the DOS are given in the next section.

We compute the effective mass of the electron using Eq.~(8) from the main part.
As a reference point that marks free space we choose the PLDOS in the middle of an idealized Fabry-P\'erot cavity with much larger mirror-separation.
The PLDOS in this case approaches the free-space value 
\begin{equation}
\rho_x \left(\omega, z = \frac{d}{2} \right) \overset{d \to \infty}{=} \frac{\omega^2}{3 \pi^2 c^3}
\end{equation}
for an arbitrary but fixed $\omega$ which is the value we use in the subtraction. 
We still include a lower cutoff of $\Lambda_{\rm IR} = 50$GHz such that we are able to assume that the effective electronic mass is given by its physical value when approximately $1$m away from the mirrors.
We point out a methodological subtlety here: While for the electron travelling close to an interface, free-space could be defined within the same system as being far away from that interface, here we compare two different systems (different boundary conditions of the electromagnetic field).
Implications of this fact will not be discussed in this work.

\subsection*{Supplement 2: Mode functions}
The quantized vector potential of the electromagnetic field is expanded in its eigenmodes according to Eq.~(1) in the main text.
The mode-functions $\vec f_q(\vec r)$ fulfill the wave-equation
\begin{equation}
    \frac{\varepsilon(\omega_q, \vec r) \omega_q^2}{c^2} \vec f_q(\vec r) - \vec \nabla \times \left( \vec \nabla \times \vec f_q(\vec r) \right) = 0
    \label{eq:WaveEquation}
\end{equation}
with suitable boundary conditions that we discuss further below.
Here $\varepsilon(\omega_q, \vec r)$ is the spatially dependent permittivity.
We use the generalized Coulomb gauge
\begin{equation}
    \vec \nabla \cdot (\varepsilon(\omega, \vec r) \vec f_q(\vec r)) = 0
    \label{eq:Gauge}
\end{equation}
which also sets the overall gauge of the vector potential.
The mode-functions additionally fulfill the normalization condition \cite{gubbin_real-space_2016}
\begin{equation}
    \int_V \mathrm{d}\vec r \, \frac{\varepsilon(\vec r, \omega_q)}{2} \left(1 + \frac{1}{\varepsilon(\vec r, \omega_q)} \frac{\partial (\omega_q \varepsilon(\vec r, \omega_q))}{\partial \omega_q}\right) |\vec f_q(\vec r)|^2 = 1
\label{eq:NormalizationDispersive}
\end{equation}
where the integral runs over the quantization volume $V$.
In the case of a non-dispersive permittivity $\varepsilon(\omega, \vec r) = \varepsilon(\vec r)$ this reduces to the well known normalization introduced in Ref.~\citenum{glauber_quantum_1991}
\begin{equation}
    \int_V \mathrm{d}\vec r \, \varepsilon(\vec r) |\vec f_q(\vec r)|^2 = 1.
\end{equation}
\paragraph*{Mode functions in the Fabry-P\'erot cavity.--}%
We describe the idealized Fabry-P\'erot cavity using perfect-conductor boundary conditions in $z$-direction at $z = 0$ and $z = d$ while using periodic boundary conditions in the $x$- and $y$-direction assuming an in-plane system size $L$ for the quantization.
In this way the distance between the mirrors of the Fabry-P\'erot cavity is set by $d$.
We set $\varepsilon(\omega, \vec r) = 1$ throughout in this case.
The electromagnetic field has two distinct polarizations: A transverse electric (TE) and transverse magnetic (TM) polarization.
The mode function for the TE case reads
\begin{equation}
    \vec f^{\rm TE}(\vec r_{||}, z) = \frac{e^{- i \vec q_{||} \cdot \vec r_{||}}}{N_{\rm TE}} \sin \left( q_z z \right) \vec e^{\rm TE}.
    \label{eq:CavityQED_fTE}
\end{equation}
Here $\vec q_{||} = (q_x, q_y)^{\rm T}$ and $\vec r_{||} = (r_x, r_y)^{\rm T}$ are the in-plane wave-vector and position-vector respectively while $q_z$ is the $z$-component of the wave-vector assuming discrete values according to
\begin{equation}
    q_z = \frac{m_z \pi}{d} \; ; \; m_z \in \mathbb{N}_0.
    \label{eq:CavityQED_ConditionKZ}
\end{equation}
The polarization vector is given as $\vec e^{\rm TE} = \frac{1}{|\vec q_{||}|} (q_y, -q_x)^{\rm T}$.
The normalization of the TE modes is given by
\begin{equation}
    N_{\rm TE}^2 = \frac{L^2 d}{2} \left( 1 - \frac{\sin(2 q_z d)}{2 q_z d} \right).
\end{equation}
The mode functions for the TM modes read
\begin{equation}
    \vec f^{\rm TM}(\vec r_{||}, z) = \frac{e^{- i \vec q_{||} \cdot \vec r_{||}}}{N_{\rm TM}} \left( \sin(q_z z) \vec e^{\rm TM}_{||} - \frac{|q_{||}|}{q_z} \cos(q_z z) \vec e_z \right)
\end{equation}
with the in-plane polarization vector $\vec e^{\rm TM}_{||} = \frac{\vec q_{||}}{|\vec q_{||}|}$ and the normalization
\begin{equation}
    N_{\rm TM}^2 = \frac{L^2 d}{2 c^2 q_z^2} \left( \omega^2 + \left(\omega^2 - 2 q_z^2 d^2 \right) \left( 1 + \frac{\sin(2 q_z d)}{2 q_z d} \right) \right).
\end{equation}
\paragraph{Mode function close to an interface.--}
We describe the interface by considering the lower half-space, $z < 0$, to be filled with a material.
The material is described by a frequency dependent but homogeneous permittivity $\varepsilon(\omega)$ such that the overall permittivity reads
\begin{equation}
    \varepsilon(\omega, \vec r) = \begin{cases}
        1 &z > 0\\
        \varepsilon(\omega) &z < 0
    \end{cases}.
\end{equation}
To describe the situation close to an insulator that hosts phonon polaritons we choose the $\varepsilon(\omega)$ as in Eq.~(10) of the main text.
In the case of an interface to a Drude metal we set $\omega_{\rm LO} \to \omega_{\rm P}$, with $\omega_{\rm P}$ the plasma frequency, and set $\omega_{\rm TO} = 0$ as described in the main part.
As before we assume periodic boundary conditions in the $x$- and $y$-direction with a finite quantization system-size of $L_{||}$.
Out of plane we once more assume perfect conductor boundary conditions at $z = \pm \frac{L_{\perp}}{2}$ in order to obtain a single interface.
However, we choose $L_{\perp}$ very large, in particular $L_{\perp} \gg \frac{c}{\omega_{\rm LO}}$, such that we are in the continuum limit practically lifting the quantization of the wave-vector in the $z$-direction. 
Unless indicated otherwise we choose $L_{\perp} = 1$m.
The electromagnetic field has both TE and TM polarization.
Taking into account continuity at the interface as well as the boundary conditions, the mode functions in the case of the TE polarization can be written as, for the upper and lower half-space respectively
\begin{equation}
    \begin{aligned}
        &\vec f^{\rm TE}(\vec r_{||}, z > 0) = \\
        &\frac{1}{N^{\rm TE}} e^{- i \vec r_{||} \cdot \vec q_{||}} \sin\left(q_z \left(\frac{L_{\perp}}{2} - z\right)\right) \sin\left(\Tilde{q}_z \frac{L_{\perp}}{2}\right) \vec e_{\rm TE}\\
        &\vec f^{\rm TE}(\vec r_{||}, z < 0) = \\
        &\frac{1}{N^{\rm TE}} e^{- i \vec r_{||} \cdot \vec q_{||}} \sin\left(\Tilde{q}_z \left(\frac{L_{\perp}}{2} + z\right)\right) \sin\left(q_z \frac{L_{\perp}}{2}\right) \vec e_{\rm TE}.
    \end{aligned}
    \label{eq:InterfaceTEModes}
\end{equation}
where $N^{\rm TE}$ denotes the normalization of the mode functions according to Eq.~(\ref{eq:NormalizationDispersive}) and one has different $z$-components of the wave-vector in the upper and lower half-space namely $q_z$ and $\Tilde{q}_z$ respectively.
They are connected through the dispersion
\begin{equation}
    \omega^2 = c^2 \left(q_{||}^2 + q_z^2 \right) = \frac{c^2}{\varepsilon(\omega)} \left(q_{||}^2 + \Tilde{q}_z^2 \right).
    \label{eq:DispersionHalfSpaces}
\end{equation}
The perfect conductor boundary conditions at $z = \pm \frac{L_{\perp}}{2}$ together with the interface conditions of the electromagnetic field only allow for a discrete set of modes.
In particular the TE modes need to fulfill the condition
\begin{equation}
    q_z \tan\left(\Tilde{q}_z \frac{L}{2}\right) + \Tilde{q}_z \tan\left(q_z \frac{L_{\perp}}{2}\right) = 0.
    \label{eq:ConditionTE}
\end{equation}

For the TM modes we split up the mode functions into in-plane and out of plane part, $\vec f = \vec f_{||} + \vec f_{\perp}$ respectively.
The in-plane part reads
\begin{equation}
    \begin{aligned}
        &\vec f_{||}^{\rm TM}(\vec r_{||}, z > 0) = \\
        &\frac{1}{N^{\rm TM}}e^{- i \vec q_{||} \cdot \vec r_{||}} \sin\left(q_z \left(\frac{L_{\perp}}{2} - z\right) \right) \sin\left(\Tilde{q}_z \frac{L_{\perp}}{2}\right) \vec e^{\rm TM}_{||}\\
        &\vec f_{||}^{\rm TM}(\vec r_{||}, z < 0) = \\
        &\frac{1}{N^{\rm TM}}e^{- i \vec q_{||} \cdot \vec r_{||}} \sin\left(\Tilde{q}_z \left(\frac{L_{\perp}}{2} + z\right) \right) \sin\left(q_z \frac{L_{\perp}}{2}\right)  \vec e^{\rm TM}_{||}.
    \end{aligned}
    \label{eq:TMModesPara}
\end{equation}
while for the out-of plane part one obtains
\begin{equation}
    \begin{aligned}
        &\vec f_{\perp}^{\rm TM}(\vec r_{||}, z > 0) = \\
        &\frac{1}{N^{\rm TM}} e^{- i \vec q_{||} \cdot \vec r_{||}} \frac{i |\vec q_{||}|}{q_z} \cos\left(q_z \left(\frac{L_{\perp}}{2} - z\right) \right) \sin\left(\Tilde{q}_z \frac{L_{\perp}}{2}\right) \vec e_z\\
        &\vec f_{\perp}^{\rm TM}(\vec r_{||}, z < 0) = \\
        &\frac{1}{N^{\rm TM}} e^{- i \vec q_{||} \cdot \vec r_{||}} \frac{-i |\vec q_{||}|}{\Tilde{q}_z} \cos\left(\Tilde{q}_z \left(\frac{L_{\perp}}{2} + z\right) \right) \sin\left(q_z \frac{L_{\perp}}{2}\right) \vec e_z.
    \end{aligned}
    \label{eq:TMModesPerp}
\end{equation}
where $N^{\rm TM}$ denotes the normalization of the mode functions according to Eq.~(\ref{eq:NormalizationDispersive}).
The z-components of the wave-vectors in both half-spaces are again related by Eq.~(\ref{eq:DispersionHalfSpaces}).
Also in the case of the TM modes, only a discrete set of modes fulfills all interface and boundary conditions.
The set is given by the solutions of
\begin{equation}
    \frac{\varepsilon(\omega)}{\Tilde{q}_z} \tan \left(q_z \frac{L_{\perp}}{2}\right) + \frac{1}{q_z} \tan\left(\Tilde{q}_z \frac{L_{\perp}}{2}\right) = 0. 
    \label{eq:ConditionTM}
\end{equation}

In general the $z$-component of the wave-vectors may be real or imaginary.
For the case of $q_z^2 > 0$ and $\Tilde{q}_z^2 > 0$ one has a travelling wave on both sides of the interface.
If $\varepsilon(\omega) > 1$ one may have $q_z^2 < 0$ in the case of a mode that is evanescent on the vacuum side (total internal reflection) and if $\varepsilon(\omega) < 1$ one may have $\Tilde{q}_z^2 < 0$ i.e.~an evanescent wave on the side of the material.
Additionally doubly evanescent waves are possible where both $q_z^2 < 0$ and $\Tilde{q}_z^2 < 0$.
These modes are exponentially localized to the interface and are thus called surface modes.
From Eq.~(\ref{eq:ConditionTE}) and Eq.~(\ref{eq:ConditionTM}) it is clear that these modes only exist in the restrahlen band when $\varepsilon(\omega) < 0$ are fully TM polarized.

We discuss the surface modes in the limit $L_{\perp} \to \infty$ writing $- i q_z = \kappa_z$ and $- i \Tilde{q}_z = \Tilde{\kappa}_z$ for imaginary $z$-components of wave-vectors on both sides of the interface.
In this case Eq.~(\ref{eq:ConditionTM}) reduces to the well known SPP condition\cite{pitarke_theory_2007}
\begin{equation}
        \kappa_z + \frac{\Tilde{\kappa}_z}{\varepsilon(\omega)} = 0.
\end{equation}
This gives a further relation between $q_z$ and $\Tilde{q_z}$ such that one may write the dispersion Eq.~(\ref{eq:DispersionHalfSpaces}) solely in terms of $\vec q_{||}$ which reads
\begin{equation}
\begin{aligned}
    &\omega^2 = c^2 q_{||}^2 \left(1 + \frac{1}{\varepsilon(\omega)}\right)\\
    & \Rightarrow \omega^2 = \frac{1}{2} \left(2 c^2 q_{||}^2 + \omega_{\rm LO}^2 - \sqrt{\omega_{\rm LO}^4 - 4 c^2 q_{||}^2 \omega_{\rm TO}^2 + 4 c^4 q_{||}^4} \right).
\end{aligned}
\end{equation}
such that one gets as the limiting frequency
\begin{equation}
    \omega^2(|\vec q_{||}|) \overset{|\vec q_{||}| \to \infty}{\to} \frac{\omega_{\rm LO}^2 + \omega_{\rm TO}^2}{2} =: \omega_{\infty}^2
\end{equation}
as stated in the main part.
The mode functions of the surface modes, in the limit $L_{\perp} \to \infty$, are given as
\begin{equation}
    \begin{aligned}
        &f^{\rm S}(\vec r_{||}, z>0) = \\
        &\frac{1}{N_{\rm S}} \left(\frac{q_x}{|\vec q_{||}|}, \frac{q_y}{|\vec q_{||}|}, i \sqrt{|\varepsilon(\omega)|} \right)^{\rm T} e^{- \frac{|\vec q_{||}|}{\sqrt{|\varepsilon(\omega)|}} z} e^{-i \vec q_{||} \cdot \vec r_{||}}\\
        &f^{\rm S}(\vec r_{||}, z<0) = \\
        &\frac{1}{N_{\rm S}} \left(\frac{q_x}{|\vec q_{||}|}, \frac{q_y}{|\vec q_{||}|}, i \frac{1}{\sqrt{|\varepsilon(\omega)|}} \right)^{\rm T} e^{\sqrt{|\varepsilon(\omega)|} |\vec q_{||}| z} e^{i \vec q_{||} \cdot \vec r_{||}}.
    \end{aligned}
    \label{eq:WFSurface}
\end{equation}
The normalization $N_{\rm S}$, as obtained via Eq.~(\ref{eq:NormalizationDispersive}), is given by
\begin{equation}
    N_{\rm S}^2 = \frac{(1 + |\varepsilon(\omega)|) \left(|\varepsilon(\omega)| + \frac{ \left[ 1 + \omega_{\rm TO}^2 \frac{\omega_{\rm LO}^2 - \omega_{\rm TO}^2}{\left(\omega^2 - \omega_{\rm TO}^2\right)^2} \right]}{|\varepsilon(\omega)|} \right)}{2 |\vec q_{||}| \sqrt{|\varepsilon(\omega)|}} L^2_{\parallel}
\end{equation}
such that, in the limit $\omega \to \omega_{\infty}$ where $\varepsilon(\omega) = -1$, the normalized mode-functions are proportional to
\begin{equation}
    f^{\rm S} \sim \sqrt{|\vec q_{||}|} e^{- |\vec q_{||}| z}
    \label{eq:App_ModeFunctionsS}
\end{equation}
as stated in the main part.
\subsection*{Supplement 3: Computation of exact local density of states}
\paragraph{Local density of states for the Fabry-P\'erot cavity.--}
Using the mode functions Eq.~(\ref{eq:InterfaceTEModes}) we compute the PLDOS parallel to the mirrors, say in $x$-direction \cite{barnes_classical_2020}.
We consider large in-plane system size $L$ such that sums over the in-plane component of the wave-vector of the field are well approximated by integrals.
We note that for consistency we formally require $L < \infty$ since otherwise two electrons might interact over infinite distances without retardation within our non-relativistic framework employing the dipole approximation.
We retain a sum for $q_z$ that remains discretized.
Starting with the TE modes, we get 
\begin{equation}
\begin{aligned}
    &\rho_x^{\rm TE}(\omega, \vec r) = \\
    & \frac{1}{2 \pi d} \sum_{q_z} \int_0^{\infty} \mathrm{d}|q_{||}| \, |q_{||}| \frac{\sin^2(q_z z)}{\left( 1 - \frac{\sin(2 q_z d)}{2 q_z d} \right)} \delta \left(\omega - c \sqrt{q_{||}^2 + q_z^2} \right). 
\end{aligned}
\label{eq:EvalDOSIntermediate}
\end{equation}
To evaluate the integral we substitute
\begin{equation}
\begin{aligned}
    &|q_{||}| \to c \sqrt{\vec q_{||}^2 + q_z^2} =: \omega'\\
    & \frac{\partial \omega'}{\partial | \vec q_{||}|} = \frac{c^2 | \vec q_{||}|}{\omega'}
\end{aligned}
\label{eq:DosSubstitution}
\end{equation}
such that we get for Eq.~(\ref{eq:EvalDOSIntermediate})
\begin{equation}
\begin{aligned}
     &\rho_x^{\rm TE}(\omega, \vec r) = \\
     &\frac{1}{2 \pi d c^2} \sum_{q_z} \int_{c q_z}^{\infty} \mathrm{d}\omega' \, \omega' \frac{\sin^2(q_z z)}{\left( 1 - \frac{\sin(2 q_z d)}{2 q_z d} \right)} \delta \left(\omega - \omega' \right).
\end{aligned}
\end{equation}
The integral vanishes if $\omega < c q_z$ and is otherwise removed by the Dirac-$\delta$.
We incorporate this condition into the sum over $q_z$ yielding the final result
\begin{equation}
    \rho_x^{\rm TE}(\omega, \vec r) = \frac{\omega}{2 \pi d c^2} \sum_{q_z, q_z < \frac{\omega}{c}} \frac{\sin^2(q_z z)}{\left( 1 - \frac{\sin(2 q_z d)}{2 q_z d} \right)}.
\end{equation}
Performing the analogous calculation for the TM modes yields
\begin{equation}
    \rho_x^{\rm TM}(\omega, \vec r) = \frac{\omega}{2 \pi d c^2} \sum_{q_z, q_z < \frac{\omega}{c}} \frac{c^2 q_z^2 \sin^2(q_z z)}{\omega^2 + (\omega^2 - 2 c^2 q_z^2) \frac{\sin(2 q_z d)}{2 q_z d}}.
\end{equation}
One may verify that when taking the limit $d \to \infty$ and a large distance to the mirror $z \gg \frac{c}{\omega}$ (remembering that the mode volume per $q_z$ is $\frac{d}{\pi}$ due to the boundary conditions) one retains
\begin{equation}
    \rho_x^{\rm TE}(\omega, \vec r) + \rho_x^{\rm TM}(\omega, \vec r) \overset{d\to \infty;\, z \gg \frac{c}{\omega}}{\to} \frac{\omega^2}{3 \pi^2 c^3}
\end{equation}
i.e.~reproducing the free-space result for a fixed but arbitrary $\omega$ fulfilling the above conditions, as expected.
\paragraph{Local density of states close to an interface.--}
In the case of the interface we compute the PLDOS in the same way as above in particular performing the substitution Eq.~(\ref{eq:DosSubstitution}).
However, through the conditions on the allowed wave-vectors Eq.~(\ref{eq:ConditionTE}) and Eq.~(\ref{eq:ConditionTM}), the $z$-component now depends on the in-plane component of the wave-vector and through it on the frequency.
Hence, in order to compute the functional determinant in the substitution, we need to perform an extra derivative
\begin{equation}
    \partial_{|\vec q_{||}|} \omega' = \frac{c^2 |\vec q_{||}|}{\omega'} + \frac{c^2 q_{z}}{\omega'} \partial_{|\vec q_{||}|} q_z.
\end{equation}
In the substitution we want to get rid of any $\vec q_{||}$ dependence, i.e.~also the derivative.
It can be replaced by a derivative with respect to the frequency employing the chain rule
\begin{equation}
    \begin{aligned}
        \partial_{|\vec q_{||}|} \omega' &= \frac{c^2}{\omega'} |\vec q_{||}| + \frac{c^2}{\omega'} q_z \partial_{\omega'} q_{z} \partial_{|\vec q_{||}|} \omega'\\
        \Rightarrow \partial_{|\vec q_{||}|} \omega' &= \frac{c^2}{\omega'} |\vec q_{||}| \left( 1 - \frac{c^2}{\omega'} q_z \partial_{\omega'} q_{z} \right)^{-1}.
    \end{aligned}
\end{equation}
Overall we can thus compute the PLDOS in the case of the interface as
\begin{equation}
    \rho(\omega, z) = \frac{L_{||}^2}{2\pi} \frac{\omega}{c^2} \sum_{\lambda} \sum_{q_z \in K(\omega)} |\vec f_{q_z, \lambda}(z)|^2 \left( 1 - \frac{c^2}{\omega'} k_z \partial_{\omega} q_{z} \right).
    \label{eq:DosInterface}
\end{equation}
where the sum runs over the set $K(\omega)$ of allowed values of $k_z$ as determined by the conditions Eq.~(\ref{eq:ConditionTE}) and Eq.~(\ref{eq:ConditionTM}) with the additional condition $q_z < \frac{\omega}{c}$.
The extra derivative $\partial_{\omega} q_z$ can be computed numerically.

\subsection*{Supplement 4: Scaling of the mass enhancement with $z^{-3}$ due to coupling to surface modes}
In the main part we have shown that the increase in effective mass close to an interface hosting surface polaritons scales as $z^{-3}$.
This scaling can be readily understood from the form of the mode functions of surface polariton modes.
First of all, we will assume that the bulk modes (all modes of the electromagnetic field except the doubly evanescent surface modes) are unaltered compared to the free-space vacuum. We have shown this assumption to hold well in Fig.~2 (c) of the main part.
We therefore cancel the contribution of the bulk-modes to the change of mass with the free-space vacuum in Eq.~(7) of the main part and conclude that for the effective change of mass close to an interface it suffices to consider the self-energy
\begin{equation}
    \Sigma^{\rm S}(\vec k_{||}) = - \alpha \frac{2 \pi \hbar^3 c |\vec k_{||}|^2}{m_e^2} \int_0^{\infty} \mathrm{d}\omega \, \frac{\rho_{\vec e_{q_{||}}}^{\rm S}(\omega, \vec r)}{\omega^2}
\end{equation}
where
\begin{equation}
    \rho_{\vec e_{q_{||}}}^{\rm S}(\omega, \vec r) = \sum_{q_{||}} |\vec f^{\rm S}_{\vec q_{||}}(\vec r) \cdot \vec e_{\vec q_{||}}|^2 \delta(\omega - \omega(\vec q_{||}))
\end{equation}
is the PLDOS due to the surface modes only.
In order to obtain the correct scaling, we consider the large $|\vec q_{||}|$ limit, $|\vec q_{||}| \gg \frac{\omega_{\rm LO}}{c}$, where we have $\omega^2 \to \omega_{\infty}^2 = \frac{\omega_{\rm LO}^2 + \omega_{\rm TO}^2}{2}$ such that we may write
\begin{equation}
    \Sigma^{\rm S}(\vec k_{||}) = - \alpha \frac{2 \pi \hbar^3 c |\vec k_{||}|^2}{m_e^2 \omega_{\infty}^2} \sum_{\vec q_{||}} |\vec f_{\vec q_{||}}(\vec r) \cdot \vec e_{\vec q_{||}}|^2.
\label{eq:App_SigmaS}
\end{equation}
For the scaling with respect to the distance to the interface $z$, we consider the sum in Eq.~(\ref{eq:App_SigmaS}) that may be written as an integral in the $L_{||} \to \infty$ limit
\begin{equation}
    \sum_{\vec q_{||}} |\vec f_{\vec q_{||}}(\vec r) \cdot \vec e_{\vec q_{||}}|^2 \to \frac{L_{||}^2}{4 \pi^2} \int \mathrm{d} \vec q_{||} \, |\vec f_{\vec q_{||}}(\vec r) \cdot \vec e_{\vec q_{||}}|^2.
\label{eq:App_SumToIntegral}
\end{equation}
The scaling as $z^{-3}$ is obtained when inserting the form of the mode function in the large $|\vec q_{||}|$ limit, Eq.~(\ref{eq:App_ModeFunctionsS}), into Eq.~(\ref{eq:App_SumToIntegral}) which yields in the large momentum limit
\begin{equation}
    \Sigma^{\rm S}(\vec k_{||}) = \frac{\hbar^2 \vec k_{||}^2}{2 m_{e}} \left(- \alpha \frac{\hbar c (\omega_{\infty}^2 - \omega_{\rm TO}^2)}{4 m_e \omega_{\infty}^4 } \frac{1}{z^3} \right).
\end{equation}
According to Eq.~(8) of the main part this directly relates to a relative change in mass due to the coupling to the surface modes that reads
\begin{equation}
    \frac{\Delta m}{m_e} = \alpha \frac{\hbar c (\omega_{\rm \infty}^2 - \omega^2_{\rm TO})}{4 m_e \omega_{\infty}^4 } \frac{1}{z^3}.
\end{equation}
which reproduces Eq.~(12) of the main part.

\bibliography{CavitronicsPaper}